\documentclass{nature}
\usepackage{graphicx}  
\makeatletter
\let\saved@includegraphics\includegraphics
\AtBeginDocument{\let\includegraphics\saved@includegraphics}

\makeatother

\usepackage{color}
\usepackage{times}
\usepackage{xcolor}
\usepackage[colorlinks=true,
            linkcolor=blue,
            citecolor=magenta]{hyperref}
\usepackage[]{amsmath}
\usepackage[]{dsfont}
\usepackage[]{amsfonts}
\usepackage[]{amssymb}
\usepackage[]{stmaryrd}
\usepackage[]{mathrsfs}
\usepackage{hyperref}
\usepackage[normalem]{ulem}
\usepackage{wasysym}

\newcommand{\ie}{i.e.,~}
\newcommand{\eg}{e.g.,~}

\usepackage{lineno}
\usepackage{xr}
\title{State-of-the-art energetic and morphological modelling of the launching
  site of the M87 jet}

\author{Alejandro Cruz-Osorio$^{1}${*}, Christian M. Fromm$^{2,1,3}${*}, Yosuke Mizuno$^{4,1}${*},
Antonios Nathanail$^{5,1}$, Ziri Younsi$^{6}$, Oliver Porth$^{7}$, Jordy Davelaar$^{8,9,10}$, 
Heino Falcke$^{10,3}$, Michael Kramer$^{3,11}$ and \\
Luciano Rezzolla$^{1,12,13}$}

\begin{document}
\maketitle

\begin{affiliations}
\small
 \item Institut f\"{u}r Theoretische Physik, Goethe Universit\"{a}t, Frankfurt, Germany
 \item Black Hole Initiative at Harvard University, Cambridge, MA, USA
 \item Max-Planck-Institut f\"ur Radioastronomie, Bonn, Germany
 \item Tsung-Dao Lee Institute and School of Physics and Astronomy,
   Shanghai Jiao Tong University, Shanghai, People's Republic of
   China
 \item Department of Physics, National and Kapodistrian University of 
   Athens, Athens, Greece
 \item Mullard Space Science Laboratory, University College London,
 Dorking, UK
\item Anton Pannekoek Institute for Astronomy, University of Amsterdam,
 Amsterdam, The Netherlands
\item Department of Astronomy and Columbia Astrophysics Laboratory,
  Columbia University, New York, NY, USA
\item Center for Computational Astrophysics, Flatiron Institute, New York, NY, USA
\item Department of Astrophysics/IMAPP, Radboud University Nijmegen,
Nijmegen, The Netherlands
 \item Jodrell Bank Centre for Astrophysics, University of Manchester,
   Manchester, UK
 \item Frankfurt Institute for Advanced Studies, Frankfurt, Germany
 \item School of Mathematics, Trinity College, Dublin, Ireland
\end{affiliations}

\noindent \textbf{M\,87 has been the target of numerous
    astronomical observations across the electromagnetic spectrum and
    Very Long Baseline Interferometry (VLBI) resolved an edge-brightened
    jet\cite{Kovalev2007,Asada2012,Walker2018,Kim2018a}.  However, the
    origin and formation of its jets remain unclear.  In our current
    understand black holes (BH) are the driving engine of jet
    formation\cite{Blandford1977}, and indeed the recent Event Horizon
    Telescope (EHT) observations revealed a ring-like structure in
    agreement with theoretical models of accretion onto a rotating Kerr
    BH\cite{EHT_M87_PaperV}. In addition to the spin of the BH being a
    potential source of energy for the launching mechanism, magnetic
    fields are believed to play a key role in the formation of
    relativistic jets\cite{Narayan2003,Narayan2012}. A priori, the spin,
    $a_\star$, of BH in M\,87$^\star$ is unknown, however, when
    accounting for the estimates on the X-ray luminosity and jet power,
    values $\left |a_\star\right| \gtrsim0.5$ appear
    favoured\cite{EHT_M87_PaperV}. Besides the properties of the
    accretion flow and the BH spin, the radiation microphysics including
    the particle distribution (thermal\cite{EHT_M87_PaperV} and
    non-thermal\cite{Kim2018b,Junor1995}) as well as the particle
    acceleration mechanism\cite{Ball2018a} play a crucial role.  We show
    that general-relativistic magnetohydrodynamics simulations and
    general-relativistic radiative transfer calculations can reproduce
    the broadband spectrum from the radio to the near-infrared regime
    and simultaneously match the observed collimation profile of M\,87,
    thus allowing us to set rough constraints on the dimensionless spin
    of M87* to be $\boldsymbol{0.5\lesssim a_{\star}\lesssim 1.0}$, with
    higher spins being possibly favoured.}

We here report the results of long-term, high-resolution,
three-dimensional (3D) general-relativistic magnetohydrodynamic (GRMHD)
simulations of magnetically arrested disks (MADs) disks around rotating BHs 
with different spin values, that is, $a_{\star}=\{- 0.9375, -0.5, 0.0, 0.5, 0.9375\}$. 
A particularly important aspect of our work is the inclusion of a non-thermal 
energy contribution, which we model in terms of a ``kappa'' distribution function
for the electron population\cite{Xiao2006,Davelaar2019}, together with a
sub-grid model that aims at modelling the effects introduced by magnetic
reconnection\cite{Mizuno21}. In this way, we
combine 3D MAD GRMHD simulations with a hybrid thermal--nont-hermal
particle distribution function to model the observed innermost structure
of the relativistic jet in M87.

\begin{figure*}[h!]
\begin{center}
\includegraphics[width=0.8\textwidth]{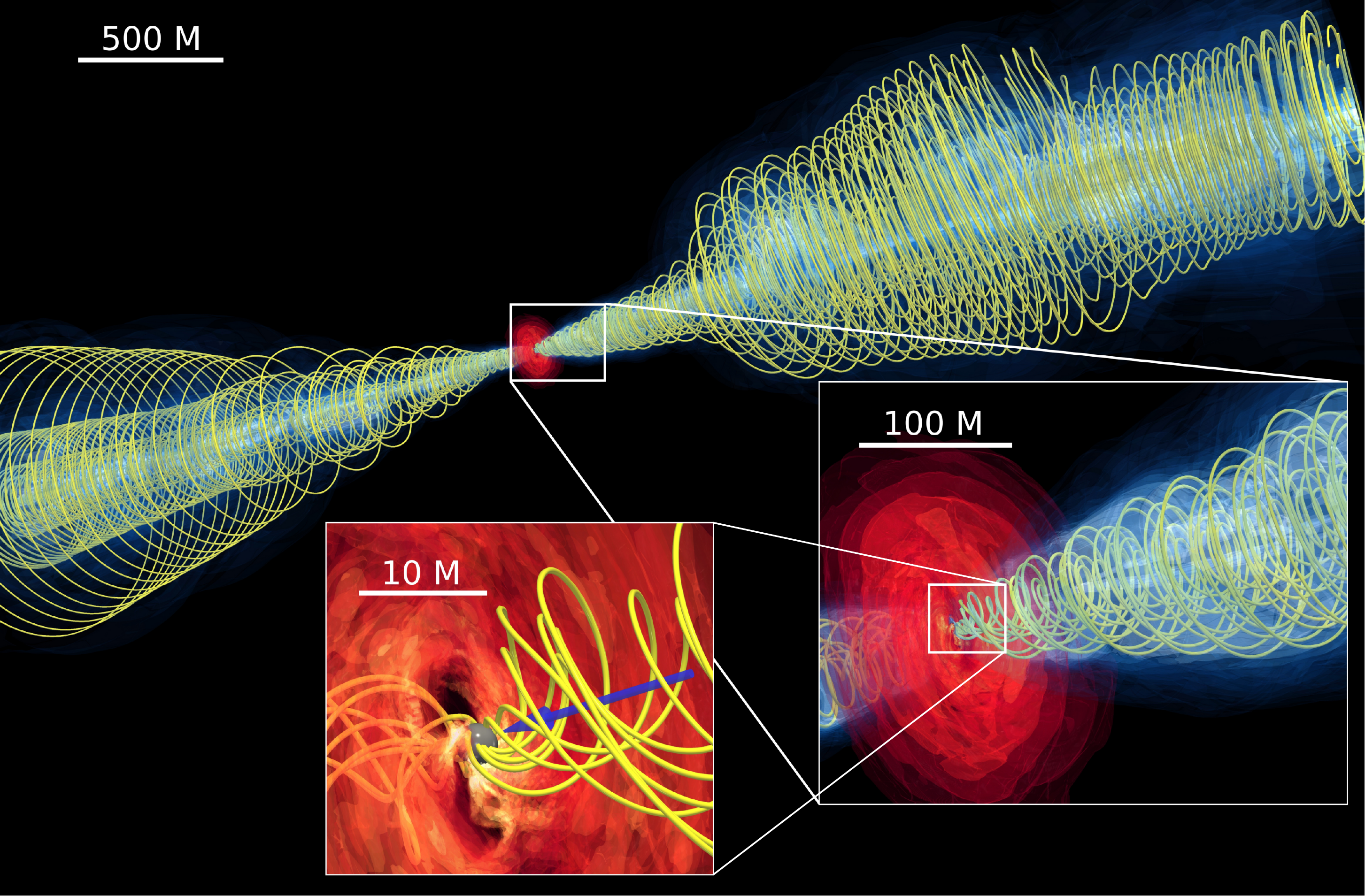}
\caption{\textbf{Large-scale 3D morphology of the jet and disk from a 3D
    GRMHD MAD simulation for a BH with $\boldsymbol{a_{\star}=0.9375}$.}
  The image shows the fast, dilute and highly magnetised out-flowing jet
  (blue) and the denser and weakly magnetised accretion disk (red).
  Magnetic field lines spiral around the jet are plotted in yellow. The
  insets offer a magnification of the large-scale image on the reported
  scales; the BH and disk angular momenta point in the bottom-left
  direction (blue arrow).}
\label{fig:3DGRMHD}
\end{center}
\end{figure*}

The first step in our modelling involves the simulation of the plasma
dynamics around the rotating BHs. To this scope, we initialise our
state-of-the-art three-dimensional GRMHD simulations with a magnetised
torus seeded with weak poloidal magnetic field in hydrodynamical
equilibrium using the numerical code \texttt{BHAC}. The development of
the magnetorotational (MRI) instability triggers accretion and leads to a
fully turbulent flow after a timescale $t=10,\!000\,M\simeq10\,{\rm yr}$,
where $M=6.5\times10^{9}\,M_{\odot}$ is the mass of M87*. At this point,
a stationary MAD solution is attained, with a the dimensionless magnetic
flux onto the BH reaches values $\Psi:=\Phi_{\rm
  BH}/\sqrt{\dot{M}}\sim15$, where $\dot{M}$ is the mass accretion rate
and $\Phi_{\rm BH}$ the magnetic flux across the horizon. Three main
regions characterise the flow: a bound, dense and weakly magnetised
accretion disk; an unbound and slow disk wind; and a
diluted, highly magnetised funnel region near the polar axis, typically
referred as the ``jet'' (see Fig.~\ref{fig:3DGRMHD} for a 3D rendering
of the simulation, and Extended Data Figures 1 and 2).

The second step requires the calculation of the radiative signatures and,
to this scope, the general-relativistic radiative transfer (GRRT)
calculations performed with the \texttt{BHOSS} code\cite{Younsi2020} need
to be tuned by fixing the mass of the BH to that of M87* and the distance to
$D=16.8\,{\rm Mpc}$. In addition, the energy distribution function of the
emitting electrons needs to be specified and the mass accretion rate
adjusted to match the observed flux densities. As anticipated, we employ
a kappa distribution function, which consists of a thermal low-energy
core and of a non-thermal high-energy tail to include the contribution of
the magnetic energy to the heating of the particles. Also, since the
GRMHD simulations model the dynamics of the non-radiating ions, a
prescription is needed to relate the temperature of the latter to the
temperature of the radiating electrons. For this, we use the so-called
``R-$\beta$'' parameterisation\cite{Moscibrodzka2014}, in which the
temperatures of the two fluids are related in terms of a free parameter
``R'' and of the ratio between the gas and magnetic pressures, \ie the
plasma $\beta$ function.

\begin{figure*}[h!]
\begin{center}
\includegraphics[width=0.9\textwidth]{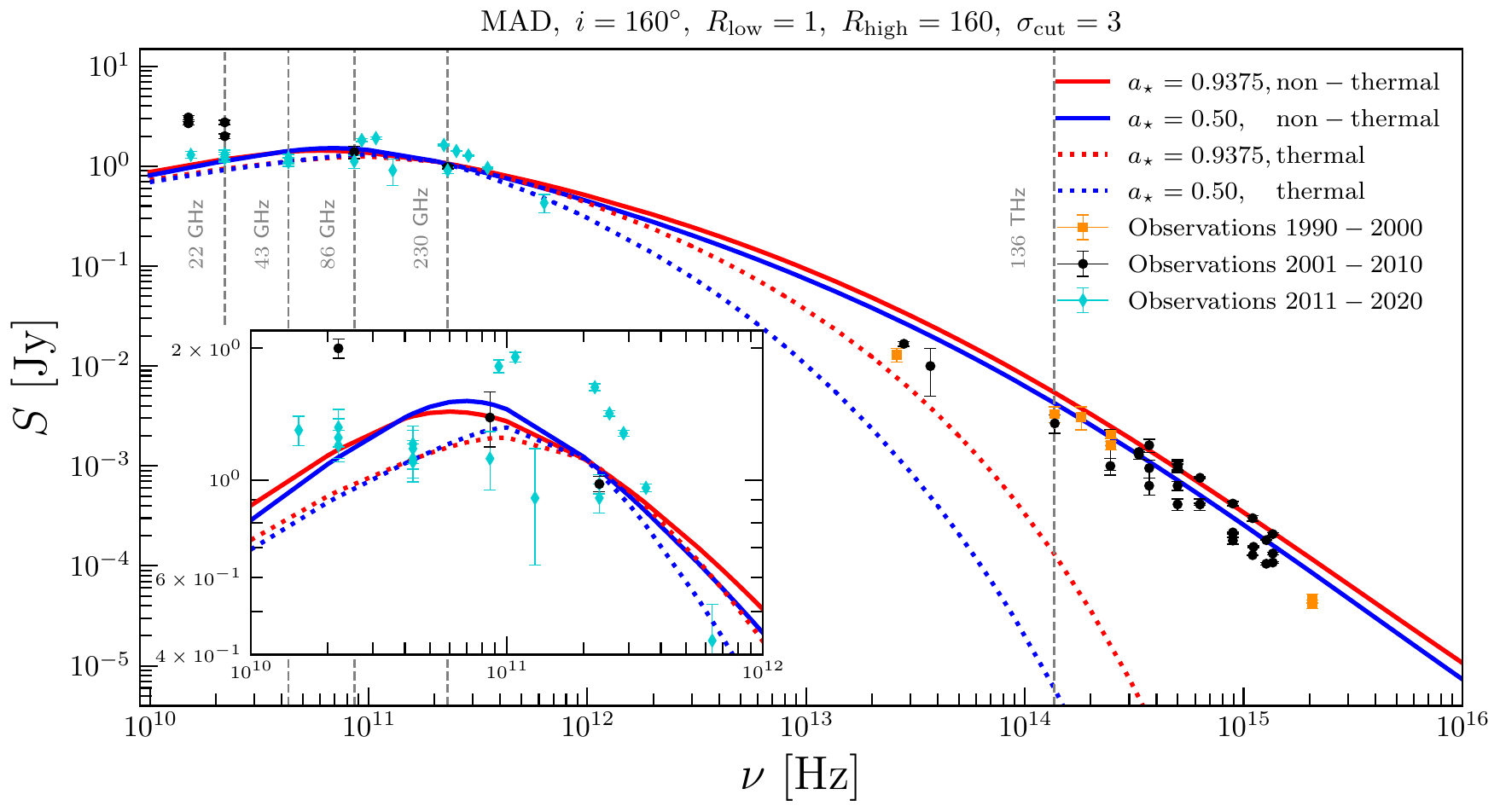}
\caption{\textbf{Broadband spectrum of the flux density of M87.} Solid
  lines show the average spectra from simulations (blue for $a_\star=0.50$
  and red for $a_\star=0.9375$) as computed between $13,\!000\,M$ and
  $15,\!000\,M$ for non-thermal models with $\epsilon=0.5$ at an inclination 
  angle of $i=160^{\circ}$, whereas dotted lines correspond to thermal models. 
  Gray vertical lines  show the most representative frequencies. The observational 
  data is indicated with symbols and covers three decades grouped into ten-years
  windows: orange squares for 1990-2000, black circles for 2001-2010, and
  turquoise-diamonds for 2011-2020 (refs. \cite{Perlman2001, Lonsdale1998,
    Perlman2001, Whysong2004, Doeleman2012, Akiyama2015, Prieto2016,
    Hada2017, An2018, Kim2018b, Lister2018, Walker2018}). 
 For each observational data, the uncertainties indicate the 
    variability during the observations.  }
\label{fig:SEDs}
\end{center}
\end{figure*}

To ensure a stable mass accretion rate and a well-developed turbulence,
we only consider the results of the simulations in the time interval
between $13,\!000\,-15,\!000\,M$ for our GRRT analysis. We scale the
mass accretion rate and adjust the additional parameters of the electron
kappa-distribution function in order to match the observed compact flux
density of $\sim1.0\,{\rm Jy}$ at 230\,GHz\cite{Doeleman2012} and to fit
the broad-band spectrum of the flux density $S$ in the range of frequencies $\nu$, 
$10^{10}\,{\rm Hz}\leq\nu\leq10^{16}\,{\rm Hz})$. The additional
parameters that need to be taken into account are: the fraction of
magnetic energy contributing to the heating of the radiating electrons,
$\epsilon$, and the injection radius of the accelerated particles,
$r_{\rm inj}=10\,M$. In Fig.~\ref{fig:SEDs} we present the results
of the spectrum obtained from the simulation that provides the best-fit
to the observed spectrum for BH spins $a_{\star}=0.5$ and
$a_{\star}=0.9375$ (for all range of BH spins see Extended Data Figure 3). 
Different symbols indicate the observational data
grouped into ten-year periods ranging from 1990 to 2020 (refs. \cite{Perlman2001,
  Lonsdale1998, Perlman2001, Whysong2004, Doeleman2012, Akiyama2015,
  Prieto2016, Hada2017, An2018, Kim2018b, Lister2018, Walker2018}).

The best-fit was obtained when using $\epsilon=0.5$ and $r_{\rm
  inj}=10\,M$, where the latter was chosen to match the location of
the stagnation surface, which is typically found between $5\,M$ and
$10\,M$ (ref. \cite{Nakamura2018}). In addition, to avoid the numerical
  contamination of our results from the highly magnetised regions of the
funnel where the GRMHD equations do not provide an accurate
  description of the plasma, we exclude regions with a very large
magnetisation, setting a cut-off at $\sigma_{\rm cut}=3$.

Clearly, Fig.~\ref{fig:SEDs} shows that the combination of MAD GRMHD
simulations together with an electron kappa-distribution function and a
magnetic reconnection sub-grid model, enables us to obtain a remarkably
good fit of the whole spectrum of M87, both in the radio ($10^{10}\,{\rm
  Hz}\leq\nu\leq10^{12}\,{\rm Hz}$) and in the near-infrared
($10^{14}\,{\rm Hz}\leq\nu\leq10^{15}\,{\rm Hz}$) bands. The novel result is 
that we have good agreement with the spectrum, accompanied by a good 
match to the jet morphology in its launching site.

\begin{figure*}[h!]
\begin{center}
\includegraphics[width=0.9\textwidth]{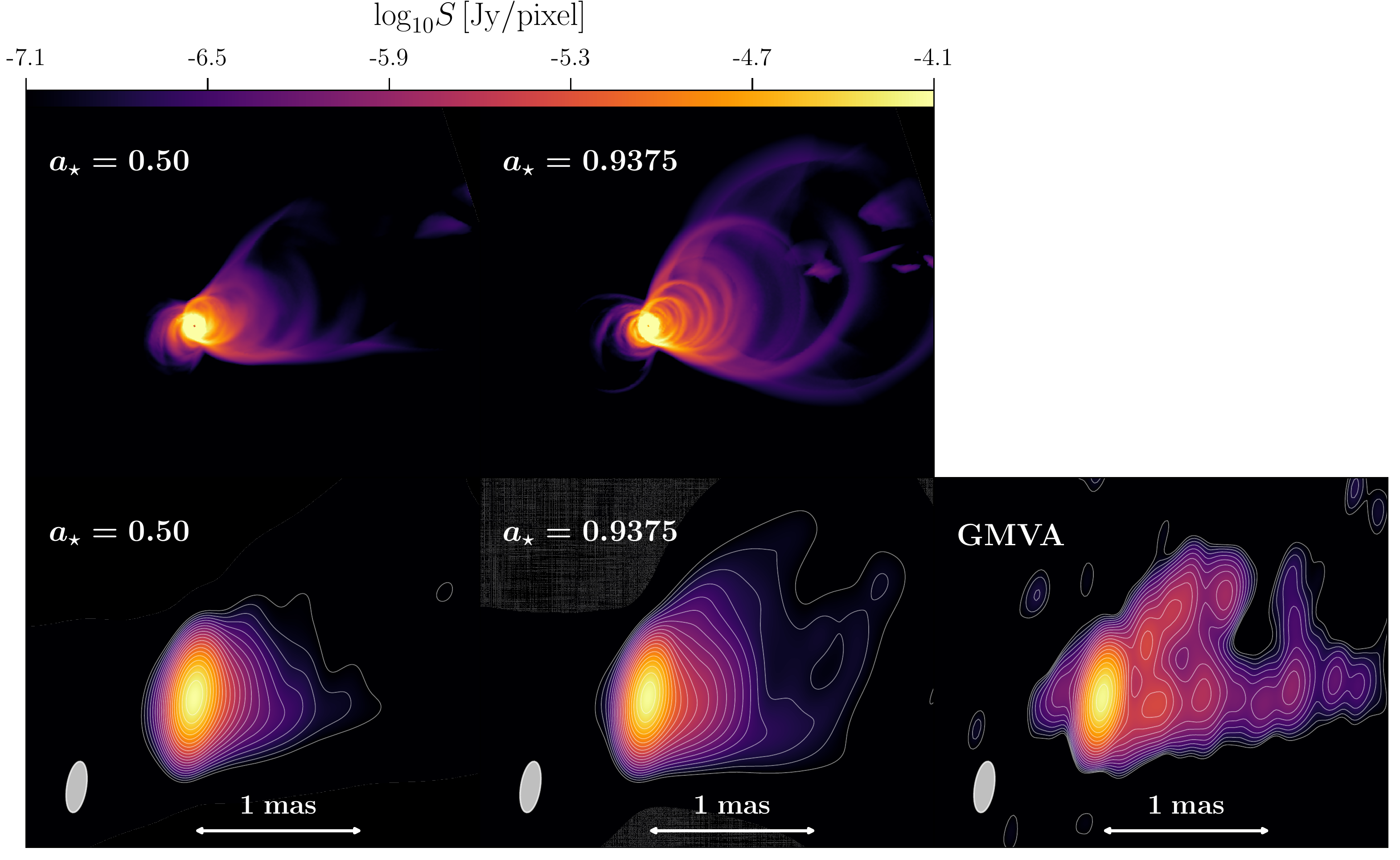}
\caption{\textbf{Morphological comparison between observations and
    theoretical models}. The top two images report the GRRT results for
  $a_\star=0.50$ (left) and $a_\star=0.9375$ (middle) at 86\,GHz. The 
  associated bottom two images show the corresponding
  convolved images,  and the image on the right offers the  GMVA
  observations. The convolving beam of $116\times307\,\mu{\rm as}$ at a
  position angle of $-9^\circ$ is shown in the lower-left corner. We
  limit the dynamical ranges of the images to three orders of magnitude,
  in agreement with current GMVA observations\label{fig:ObsandTeo}.}
\end{center}
\end{figure*}

In addition to the broad-band spectrum, our simulations can
  reproduce the morphology of the M87 jet with high accuracy up to
  distances $r \sim 1\ {\rm mas}$ from the core, as deduced from the
  Global mm Very Long Baseline Interferometry Array (GMVA)  observation
  in February 2014 \cite{Kim2018b}. To perform such a
comparison, we generate 200 GRRT images covering a time span of
$2,\!000\,M$ at 86\,GHz (3\,mm). Furthermore, we limit the dynamical
range of the flux of our synthetic images to be within three orders of
magnitude from the maximum flux, \ie $\log_{10} S \in [-7.1,-4.1]\,{\rm
  Jy/pixel}$ in agreement with the GMVA observations. In
Fig. \ref{fig:ObsandTeo} we present the best-fit images -- as deduced
from the normalised cross-correlation coefficient and
structural-dissimilarity index\cite{Mizuno18} computed between the GRRT
images and the GMVA observations. The top  two images show the
high-resolution simulations, whereas the associated bottom two images 
show the result of the convolve of the simulations with a beam of 
$116\times307\,\mu{\rm as}$ to mimic the limited resolution of 
the GMVA observations. The 2014 GMVA image is shown on the right. 
 Note that both models produce large opening angles, 
but also that the characteristic ``limb brightening'' of the 2014 GMVA image is
visible and closely similar only for the $a_\star=0.9375$
model. Furthermore, the rapidly spinning BH also displays a detectable
signature of a counter jet, again in good in agreement with the 2014 GMVA
observations (Extended Data Figures 4 and 5 also show thermal 
model and averaged images).

\begin{figure*}[h!]
\begin{center}
\includegraphics[width=0.65\textwidth]{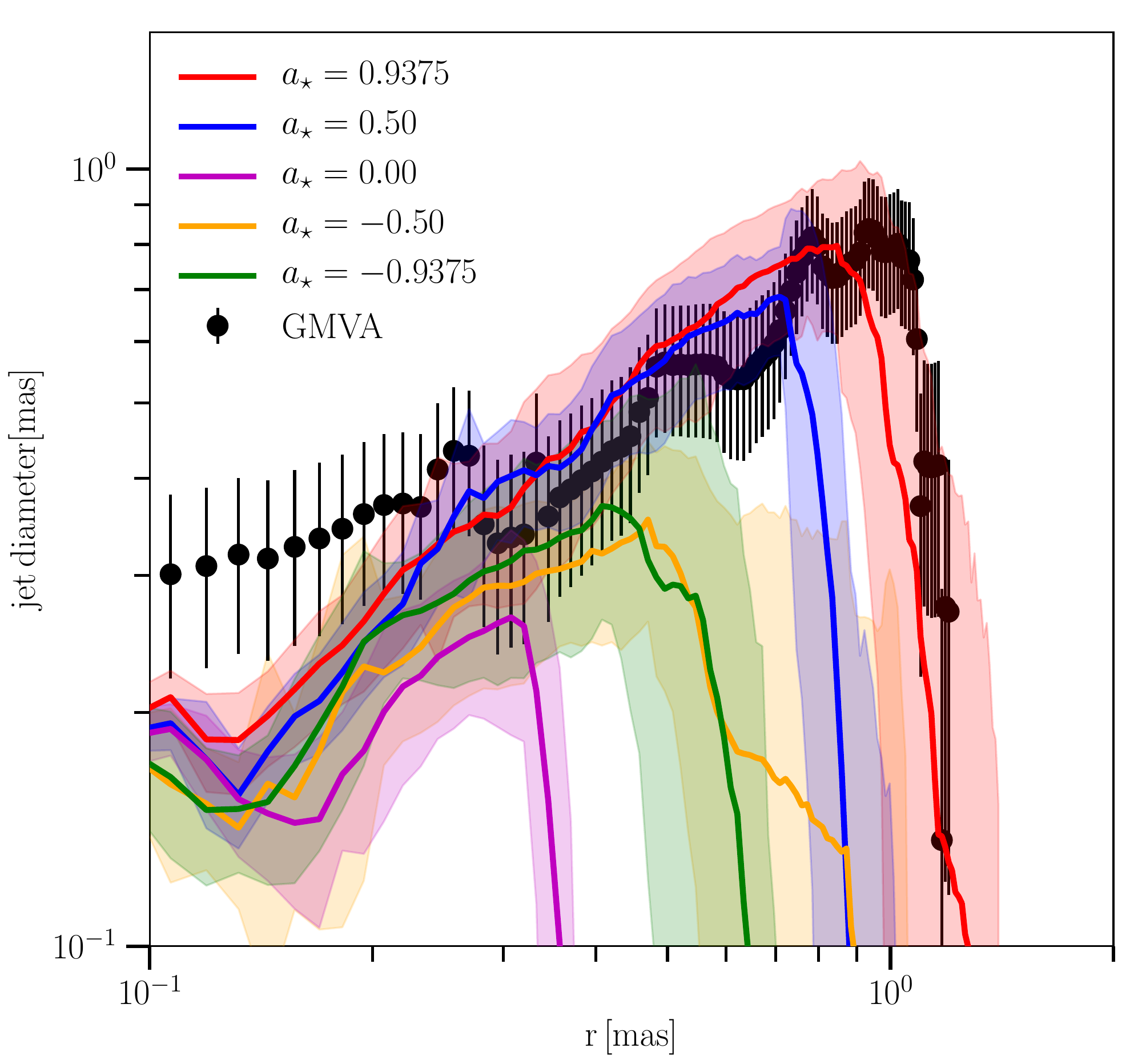}
\caption{\textbf{Jet-diameter comparison between observations and
    theoretical models}. Shown with black points is the jet width
  computed from the GMVA observation where the uncertainties are obtained
  by assuming an uncertainty of $1/4$ of the beam size at the $r=0$ and a
  linear increase until $1/2$ of the beam size is reached at $r=2\,{\rm
    mas}$ (ref. \cite{Mertens2016}). Reported instead with solid lines are the
  equivalent measures from the convolved GRRT images covering a time span
  of $2,\!000\,M$ (green for $a_\star=-0.9375$, orange for $a_\star=-
  0.50$, magenta for $a_\star=0.00$, red for $a_\star=0.50$ and blue for
  $a_\star=0.9375$); shown as shaded regions are variations within the
  standard deviation. }
\label{fig:jetwidth}
\end{center}
\end{figure*}

A more quantitative measure of the properties of the jet structure as
obtained from the observations and from simulations can be
obtained by measuring the jet diameter. For this scope, we slice the jet
in a direction orthogonal to that of propagation and fit the flux-density
profile using up to three different Gaussians. The jet width is then
computed as the distance between two outermost Gaussians and this shown
in Fig. \ref{fig:jetwidth} for various spins. Note that at small
distances, \ie for $r<0.2\,{\rm mas}$, the two models with different
spins yield jet widths that are smaller than the one deduced from GMVA
observations of M87. However, for larger distances, \ie for $0.2\,{\rm
  mas}\leq r\leq 0.7\,{\rm mas}$, the jet widths are in good agreement
with the observations for both spins; this morphological match is lost
for simulations with $a_\star\lesssim0.5$ ( see Extended Data 
Figure 6 for a quantitative measurement).

In summary, the results presented here clearly show that long-term,
high-resolution and state-of-the-art GRMHD simulations of magnetically
arrested disks in combination with a non-thermal description of the
electron energy distribution can reproduce not only the broad-band
spectrum, but also the 86\,GHz emission structure of M87. Interestingly,
if non-thermal particles are included in the GRRT calculations, both
models are able to reproduce the observed spectrum of M87, including a
flat spectrum in the radio ($10^{10}\,{\rm Hz}\leq\nu\leq10^{12}\,{\rm
  Hz}$) and in the near-infrared ($10^{14}\,{\rm Hz} \leq \nu\leq
10^{15}\,{\rm Hz}$) bands. Furthermore, since both MAD models naturally produce
jets with large opening angles, the simulations provide a remarkably good
match of the morphology of the jet as measured in terms of the diameter
along the direction of propagation. Because this match is lost for BHs
with small or negative spins, our results constrain the dimensionless
spin of M87* to be roughly in the range $0.5\lesssim
a_{\star}\lesssim1.0$, with higher spins being possibly favoured.

The promising results presented here call for additional work and
improvements, where modelling including particle heating, radiative
cooling and polarisation will further improve the ability of our models
to reproduce also finer details of the observed features of the M87
jet. In addition, the ongoing technical improvements of the Very Long Baseline Interferometry
technique will allow us to image the jet footpoint with unprecedented
resolution and further constrain our models.

\begin{methods}

\section*{General-relativistic magnetohydrodynamic simulations of
    magnetically arrested disks.}

The simulations presented in the main text were performed using the
state-of-the-art multidimensional GRMHD code \texttt{BHAC v1.1}  \cite{Porth2017}, 
which has been designed
to solve the equations of GRMHD in arbitrary but fixed spacetimes and
different coordinate systems; the results presented here refer to 3D
simulations using spherical polar coordinates. The numerical techniques
are based on second-order finite-volume and high-resolution
shock-capturing methods\cite{Rezzolla_book:2013}, complemented with
flux-constrained transport of the magnetic field so as to preserve its
divergence-free constraint $\nabla \cdot {\boldsymbol {\rm 
    B}}=0$ (ref. \cite{Olivares2018a}). We have carried out simulations
considering that the spacetime is described by the Kerr solution written
in Kerr-Schild coordinates with dimensionless spin parameter
$a_{\star}=\{- 0.9375, -0.5, 0.0, 0.5, 0.9375\}$. Most of our attention,
however, is concentrated on the cases of positively and rapidly spinning
BHs, \ie $a_{\star}=0.5$ and $a_{\star}=0.9375$, as these are the
configurations that provide the best morphological match to the GMVA
observations.

As initial data for our 3D GRMHD simulations, we consider a magnetised
torus with constant specific angular momentum, $\ell:=u_\phi/u_t =
6.8\ (a_{\star}=0.5)$ and $6.76\ (a_{\star}=0.9375)$ in hydrodynamic
equilibrium around the rotating BH, where $\boldsymbol{u}$ is the fluid
four-velocity. A weak, single-loop poloidal magnetic field is added on
the hydrodynamical solution in terms of the vector potential
$A_{\phi}\propto\mathrm{max}(q-0.2,0)$, where $q:=(\rho/\rho_{\rm
  max})\left(r/r_{\rm in}\right)^3\sin^3\theta\exp\left(-r/400\right)$,
and normalised so that
$\beta_{\mathrm{min}}=(2p/b^2)_{\mathrm{min}}=100$, where the subscript
$\mathrm{min}$ refers to the minimum value inside the torus. The location
of the inner edge and central radius of the torus, defined by the
equipotential surfaces of the gravitational-centrifugal potential, are
$r_{\rm in}=20\,M$ and $r_{\rm centre}=40\,M$,
respectively\cite{Font02b,Rezzolla_book:2013}.
We make use of an ideal-gas equation of state\cite{Rezzolla_book:2013}
with an adiabatic index of $\Gamma=4/3$ to model the relativistic gas in
the torus. As is customary in this type of simulations, we excite the magnetorotational instability
in the torus by adding a $4\%$ random perturbation to the equilibrium gas
pressure of the torus. To avoid numerical failures due to the presence of
vacuum regions outside the torus, it is conventional to add an
``atmosphere'' to regions which should be virtually in vacuum. For all
numerical cells whose rest-mass density satisfies the condition $\rho \le
\rho_{\rm fl}$, or whose gas pressure is such that $p \le p_{\rm fl}$, we
simply set $\rho=\rho_{\rm fl}$ and $p = p_{\rm fl}$, where the floor
values are $\rho_{\rm fl}:=10^{-4}\,r^{-3/2}$ and $p_{\rm
  fl}:=(10^{-6}/3)\,r^{-5/2}$ and are based on the exact solution of the
Bondi-accretion problem.

The spherical polar (spatial) coordinate map $(r,\theta,\phi)$ covers the
simulation domain in the range $r\in[0.8\,r_{\mathrm{EH}},2500\,M]$,
$\theta\in[0,\pi]$, and $\phi\in[0,2\pi]$, where
$r_{\mathrm{EH}}=M+\sqrt{M-a_{\star}^2}$ is the event-horizon radius, and
with the innermost cell well inside the event horizon. The grid spacing
is logarithmic in the radial direction and uniform in the two angular
$\theta-$ and $\phi-$ directions, using three refinement levels with the
effective number of grid points given by
$(N_r,N_\theta,N_\phi)=(384,192,192)$. At the inner radial boundary,
which is within the event horizon thanks to the horizon-penetrating
coordinates, we apply an inflow boundary condition, while at the outer
radial boundary we enforce an outflow boundary condition by performing a
copy of the physical variables. At the polar boundaries, \ie for
$\theta=0$ and $\theta=\pi$ we apply the so-called ``hard'' polar
boundary conditions, namely, we assume that a solid reflective wall is
present along the polar boundaries, where the flux and poloidal velocity
through the boundaries is set to be zero, adjusting the electromotive
force in the constrained transport
routine\cite{Shiokawa2012,Olivares2018a}. For the azimuthal direction,
instead, periodic boundary conditions are employed in all physical
variables across the cells at $\phi=0$.

\section*{General-relativistic radiation-transfer with
  a non-thermal emission model.}

In order to compute the radio images of the accretion flow around M87*
and of the corresponding jet, we have employed the GRRT code
\texttt{BHOSS}\cite{Younsi2020,Gold2020}. In this code, electromagnetic
radiation propagates along null geodesics, which are used to solve the
radiative-transfer equation along such path\cite{Younsi2012}. To
generate large number of images over a considerable time range, we
consider for all simulations the time interval
$t\in[13,\!000\,M-15,\!000\,M]$, with a cadence of $10\,M$; the resulting
200 images cover about two years of observations given the
estimated mass of M87*, which we take to be $6.5
\times10^9\,\mathrm{M}_{\odot}$ (ref. \cite{EHT_M87_PaperI}). The image size is
then also computed adopting a distance from M87* of
$16.8\,{\rm Mpc}$ (ref. \cite{EHT_M87_PaperI}).

The electron temperature in the jet is computed through the temperature
of the ions in the so-called ``two-temperatures'' model so that the
ion-to-electron temperature ratio is expressed as $T_{\rm i}/T_{\rm
  e}:=(R_{\rm low}+R_{\rm
  high}\beta^{2})/(1+\beta^{2})$ ( ref. \cite{Moscibrodzka2009}), which depends on
the ratio between the gas and magnetic pressures, \ie the plasma $\beta$
function, which is large inside the accretion flow and small
elsewhere. The relation between the two temperatures also makes use of
two free coefficients, $R_{\rm low}$ and $R_{\rm high}$, whose values for
regions involving the emission from the jet ($\beta\ll1$), and from the
disk ($\beta\gg1$), have been set to be $R_{\rm low}=1$ and $R_{\rm
  high}=160$, respectively. These values are routinely adopted in the
two-temperatures model and we have systematically verified that they also
provide the best match to the broadband spectrum and to the jet
morphology.

An essential aspect of our GRRT modelling consists in the use of a
non-thermal energy distribution for the electrons that we model, following
Davelaar et al. (2019)\cite{Davelaar2019}, in terms of the so-called
''kappa'' distribution, which is a combination of a thermal electron
population and of a population with a power-law energy distribution, \ie
$d n_{\rm e} /d \gamma_{\rm e} = N \gamma_{\rm e} \sqrt{ \gamma_{\rm
    e}^{2}-1} \left[ 1 + (\gamma_{\rm e}-1)/(\kappa w)
  \right]^{-(\kappa+1)}$, where $\gamma_{\rm e}$ is the Lorentz
factor of the electrons and $N$ is a normalization
parameter. From a more phenomenological point of view, the so-called
kappa-jet model intends to describe the energy contribution of electrons
accelerated due to magnetic reconnection at the base of the jet. The
weighted temperature $w$ is defined as $w:=(\kappa-3)\Theta_{\rm
  e}/\kappa+\tfrac{1}{2}\epsilon\left[1+\tanh{(r-r_{\rm
      inj})}\right](\kappa-3)m_{\rm p}\sigma/(6 \kappa\ m_{\rm e})$,
where $r_{\rm inj}=10\,M$ is the injection position, $\Theta_{e}$ is the dimensionless electron temperature, 
$m_{\rm p}$ and
$m_{\rm e}$ are respectively the proton and electron masses, $\sigma$ is
the magnetisation parameter, \ie the ratio of the magnetic and
rest-mass energy densities, and $\epsilon$ is a tunable parameter between
zero and one that accounts for the fraction of magnetic energy
contributing to the heating of the radiating electrons. Finally, the
power-law index $\kappa$, depends on the microphysics of the plasma and
the expression employed here, namely 
$\kappa:=2.8+0.7\,\sigma^{-1/2}+3.7\,\sigma^{-19/100}\tanh{(23.4\,\sigma^{26/100}
  \beta)}$, is inspired by particle-in-cell
simulations\cite{Ball2018a,Davelaar2019} and is set to be a function of
magnetisation parameters $\beta$ and $\sigma$ only.

Extended Data Figures 1 and 2 report the
large-scale morphology of the jet from GRMHD simulations of a Kerr BH
with spin $\mathbf{a_{\star}=0.9375}$. Extended Data Figure 1
concentrates on the large-scale structure of the jet, whereas 
Extended Data Figure 2 offers a magnification near the event
horizon. In both cases, shown from left to right are the magnetisation
parameters $\sigma$, the electron temperature $T_{\rm e}$, the
distribution of the power-law index $\kappa$, and the weighted
temperature $w$. All quantities are averaged in space (over the azimuthal
direction) and in time (over a time interval of $2,\!000\,M$ with a
cadence of $10\,M$). The white dashed line marks the boundary between
bounded (\ie matter with Bernoulli parameter ${\rm Be}:=-hu_t<1.02$,
where $h$ is the specific enthalpy\cite{Rezzolla_book:2013}) and
unbounded (${\rm Be}>1.02$) material, whereas the black lines show the most
important contours of the logarithm of the magnetisation. Also reported
away from the polar axis are the contours corresponding to $\sigma=3.0$,
which we take to separate the ``jet spine'' (defined as $\sigma >
\sigma_{\rm max}:=3.0$), from the ``jet sheath'', (for which $0.1 <
\sigma < 3.0$), and from the ``external wall'' (defined as $\sigma=0.1$).

To compute the mass accretion rate, we normalise the emission at a
resolution of $800\times800$ pixels -- which corresponds to a field of
view of $4\, {\rm mas}\ \sim 10^3\,M$ -- so as to reproduce the observed
flux of M87 at 230\,GHz of $\simeq 1.0\,{\rm Jy}$ (refs. \cite{Akiyama2015,
  Doeleman2012}), excluding the emission from the jet-spine region, \ie
regions with $\sigma>3.0$. The resulting mass accretion rates are
$\dot{M}=3.33\times 10^{-4}\, M_{\odot}\,\mathrm{yr}^{-1}$ and
$\dot{M}=1.06\times 10^{-4}\, M_{\odot}\,\mathrm{yr}^{-1}$ for the two
rapidly rotating BHs with $a_{\star}=0.5$ and $a_{\star}=0.9375$,
respectively. The inclination angle of the observation is assumed to be
fixed at $160^{\circ}$, as deduced from the recent observations of the
EHT\cite{EHT_M87_PaperV}.

To constrain our numerical simulations and extract information
on the free parameters of our models, we have used historical
observations of the broadband spectrum of M87 as well as the jet
morphology in terms of the jet opening angle and jet width measured from
the core and along the direction of propagation. After having fixed the
BH spin and the jet-spine wall, \ie after having set a value for
$\sigma_{\rm cut}$, we have explored the electron-temperature parameters
by varying the values of the free coefficients $R_{\rm low}$ and $R_{\rm
  high}$ as a way to heat or cool the electrons in the jet funnel. In
this way, we have found that large-scale, coherent jet morphology can be
imaged only for large values of $R_{\rm high}$, \eg $R_{\rm high}=160$,
quite independently of the electron distribution function, \ie whether a
non-thermal emission is included or not.

However, although not important for the morphology, the
inclusion of a non-thermal component is important to obtain a good match
with the observed spectrum and in the near-infrared, in
particular. Hence, after finding the optimal description for the electron
temperature with $R_{\rm low}=1$ and $R_{\rm high}=160$, we have explored
the magnetic-energy contribution in the non-thermal component by varying
the parameter $\epsilon$ for fixed values of the BH spin and of
$\sigma_{\rm cut}$. In this way, we found good a agreement with jet
morphology for $0.25 \lesssim \epsilon \lesssim 1$, noting, however that
the near-infrared flux is over-produced with respect to the observations
$ \epsilon \geq 0.75$. At the same time, we have explored different
locations of the jet-spine wall by changing the cut-off value for the
magnetisation $\sigma_{\rm cut}$, see also Chael et al. (2019)
\cite{Chael2019}. In this way, it was possible to
determine that the highly magnetised region in the jet spine
contributes substantially to the near-infrared emission, so that a
combined good match of the observed broadband spectrum and of the
morphology can be obtained for $3 \lesssim \sigma_{\rm cut}
  \lesssim 5$. We also performed the GRRT calculations for different BH
spins fixing the electron temperature parameters, the location of the
jet-spine wall, and the parameters of the non-thermal energy contribution,
as discussed above. In this way we found only small variations in the
radio and near-infrared bands when changing the spin of the BH from
co-rotating to counter-rotating cases. The average thermal and non-thermal
spectra for the best-fit parameters have been presented in
Extended Data Figure 3 for the whole range of BH spins considered
here. Moreover, because small and counter-rotating BHs show small opening
angles and very short jets, a good agreement with the observed jet
morphology can be achieved only for $0.5\lesssim
a_{\star}\lesssim1.0$. This is shown in Extended Data Figure 4 that reports
the GRRT images (top row) and the convolved ones (bottom row) for thermal
and non-thermal emission models, and for two BHs with spins
$a_{\star}=0.50$ and $a_{\star}=0.9375$, respectively. The images refer
to a representative time $t=13,\!820\,M$, and we show as an ellipse in
the lower-left corner the convolving beam with axes
$116\times307\,\mu{\rm as}$, as in the observational data. To remark that
these features are not dependent on the specific choice of the image
frame, we show in Extended Data Figure 5 the same same quantities as in
Extended Data Figure 4 but after performing and average in time between
$13,\!000\,M$ and $15,\!000\,M$. Clearly, all of the features discussed
in Extended Data Figure 4 for a single snapshot are present also when
averaging in time; a more detailed discussion of this point will be
presented elsewhere.

In summary, when combining all the constraints deduced for the electron
temperature, the jet spine, the magnetic energy in the non-thermal model,
and the BH spin, we found that the best-fit model obtained from the
numerical simulations is the one obtained for $R_{\rm low} =1$, $R_{\rm
  high}=160$, $\sigma_{\rm cut}=3$, $\epsilon=0.5$ and $0.5\lesssim
a_{\star}\lesssim1.0$, respectively. When comparing our results
  with those already published in the literature to constrain the spin of
  M87*, we find that there is an overall good agreement, despite the use
  of different techniques. In particular, some authors find that
  $a_{\star} \geq 0.4$ is needed to generate a sufficiently large
  magnetic flux to launch the jet\cite{Nemmen2019}, whereas a higher spin
  value, \ie $a_{\star} \sim 0.98$, was deduced when using only the spectral energy distribution
  of M87* observations, in particular the jet power in the X-ray band, to
  constrain a jet within an advection-dominated accretion flow (ADAF)
  model\cite{Feng2017}. Finally, our results are also in agreement with
  the estimates derived by the EHT when accounting for the X-ray
  luminosity and jet power ($\left |a_\star\right|
  \gtrsim0.5$) (ref. \cite{EHT_M87_PaperV}), and with the very recent polarised
  images at 230\,GHz recently presented from EHT observations of M87*,
  where high spins of  $0.5\lesssim
  a_{\star}\lesssim1.0$ were favoured \cite{EHT_M87_PaperVII,EHT_M87_PaperVIII}.

\section*{Convolved imaging and comparison with the observations}

To reproduce the GMVA observations, we have convolved the
ray-traced images at 86\,GHz with an elliptical beam of
$307\times116\,\mu{\rm as}$ at a position angle of $-9^\circ$, which
corresponds to the March 2014 GMVA observation of M87 (ref. \cite{Kim2018a}).
To obtain the best-fit models presented in the lower row of
Fig. \ref{fig:ObsandTeo}, we compute the normalised cross-correlation
coefficient and the structural dis-similarity image measure (DSSIM)
between the GMVA image and the convolved GRRT images\cite{Mizuno18}.

Furthermore, to compute the jet width, we sliced the jet transversal to
the jet axis (defined the direction of propagation) and fitted the flux
density profile with two or three different Gaussian distributions. The
third Gaussian is typically required to produce a good fit, \ie
$\chi^2\sim 1$ (see Extended Data Figure 6). The final jet width is then
measured between the peaks of the two outermost Gaussians\cite{Kim2018b};
the error estimate for the jet width from the GMVA observations, 
by contrast, is calculated following the approach of Mertens et
al.\cite{Mertens2016}. For the convolved GRRT images, we computed the
mean and standard deviation of the jet width by applying the above
described method to our entire data set (see Fig. \ref{fig:jetwidth}).

\end{methods}
\begin{addendum}

\item[Acknowledgements] This research is supported by the ERC synergy
  grant ``BlackHoleCam: Imaging the Event Horizon of Black Holes'' (Grant
  No. 610058). CMF is supported by the Black Hole Initiative at Harvard
  University, which is supported by a grant from the John Templeton
  Foundation. AN was supported by the Hellenic Foundation for Research
  and Innovation (H.F.R.I.) under the ``2nd Call for H.F.R.I. Research
  Projects to support Post-Doctoral Researchers'' (Project Number:
  00634). ZY is supported by a UK Research and Innovation Stephen Hawking Fellowship and
  acknowledges support from a Leverhulme Trust Early Career Fellowship.
  JD is supported by NASA grant NNX17AL82 and a Joint Columbia/Flatiron
  Postdoctoral Fellowship. Research at the Flatiron Institute is
  supported by the Simons Foundation. The simulations were performed on
  Goethe-HLR at the Scientific Computing in Frankfurt, on Iboga at the Institut 
  f\"ur Theoretische Physik, and Pi2.0 at Shanghai Jiao Tong University.
\item[Author Contributions] ACO and CMF performed and analysed the GRRT
  calculations and wrote the manuscript. YM performed the GRMHD
  simulations and wrote the manuscript. AN wrote the manuscript. ZY
  authored the GRRT code \texttt{BHOSS}. OP authored the GRMHD code
  \texttt{BHAC}. JD helped in GRRT calculations.  H. F. and M. K. wrote the 
  manuscript. LR wrote the manuscript
  and coordinated the various aspects of the research. All authors
  discussed the results and commented on all versions of the manuscript.
 \item[Correspondence] Correspondence and requests for materials should
   be addressed to Alejandro Cruz-Osorio~(email: 
   osorio@itp.uni-frankfurt.de), Christian
   M. Fromm~(email:cfromm@itp.uni-frankfurt.de) and 
   Yosuke Mizuno ~(email:mizuno@sjtu.edu.cn).
\item[Data availability] The data that support the plots within
  this paper and other findings of this study are available from the
  corresponding authors upon reasonable request.
\item[Code availability] The public released version of the GRMHD 
code \texttt{BHAC} can be found at \href{https://bhac.science}{https://bhac.science}. 
The  \texttt{eht-imaging} software to convolve the images is available at  repository
\href{https://github.com/achael/eht-imaging}{https://github.com/achael/eht-imaging}.
 \item[Competing Interests] The authors declare that they have no
   competing financial interests.
\end{addendum}
\renewcommand{\figurename}{Extended Data Figure}
\setcounter{figure}{0}
\begin{figure*}
\begin{center}
\includegraphics[width=0.99\textwidth]{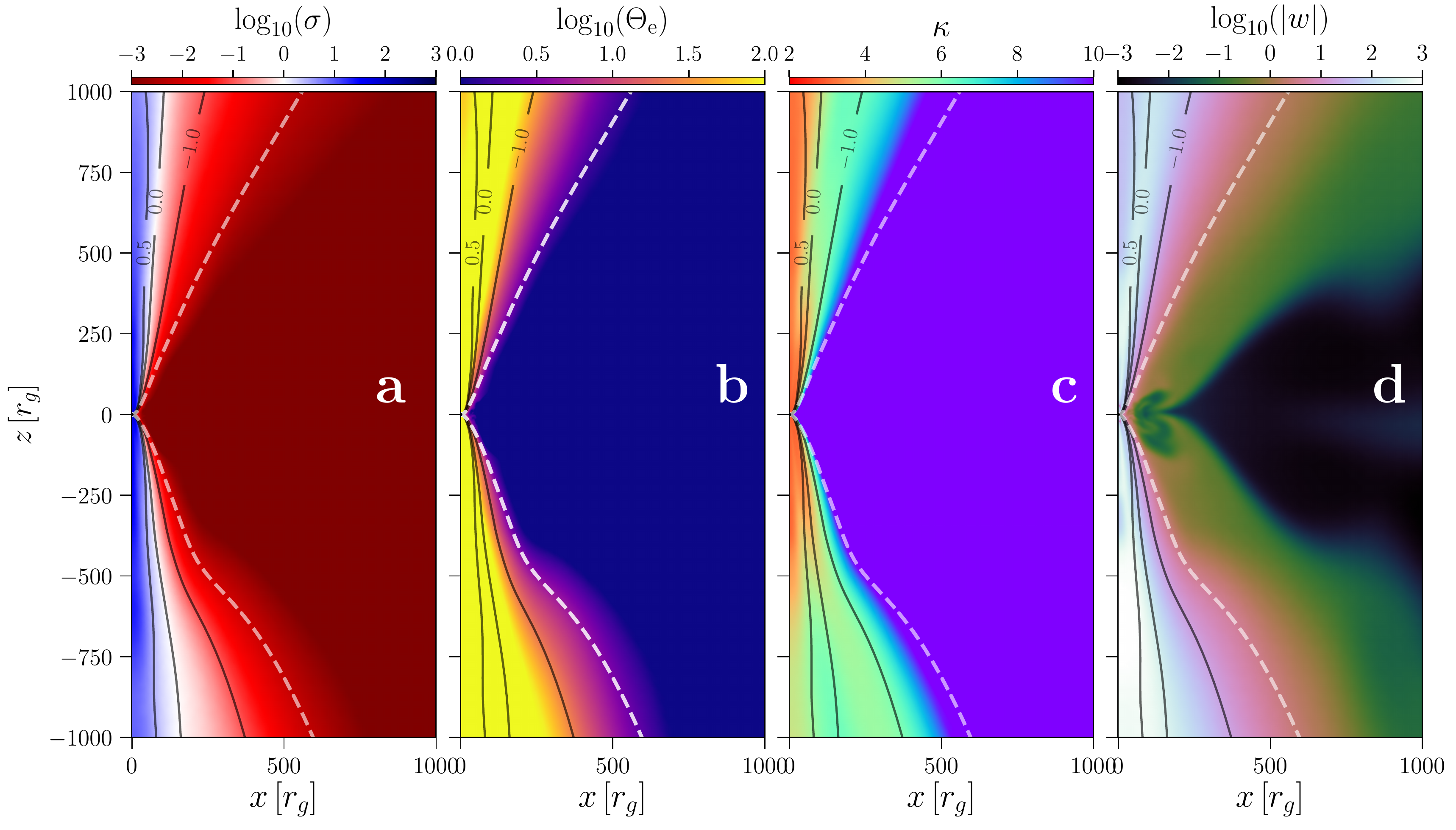}
\caption{\textbf{Large-scale morphology of the
    jet from GRMHD simulations of a Kerr BH with spin
    $\mathbf{a_{\star}=0.9375}$ (the BH spin is aligned with the $z$
    axis).} \textbf{a)} Shown the magnetisation parameters
  $\sigma$, \textbf{b)} the electron temperature $\Theta_{\rm e}$, \textbf{c)} the distribution of the
  power-law index $\kappa$, and \textbf{d)} the weighted temperature $w$. All
  quantities are averaged in space (over the azimuthal direction) and in
  time (over a time interval of $2,\!000\,M$ with a cadence of
  $10\,M$). The white dashed line marks the boundary between bounded and
  unbounded (${\rm Be}>1.02$) material, while the black lines show the
  most important contours of the logarithm of the magnetisation,
  $ \log_{10} \sigma =-1.0,\ 0.0$, and $0.5$. Moving
  out from the polar axis, we report the contours for the ``jet spine'',
  the ``jet sheath'', and the ``external wall''.}
\label{fig:grmhd_large}
\end{center}
\end{figure*}

\begin{figure*}
\begin{center}
\includegraphics[width=0.99\textwidth]{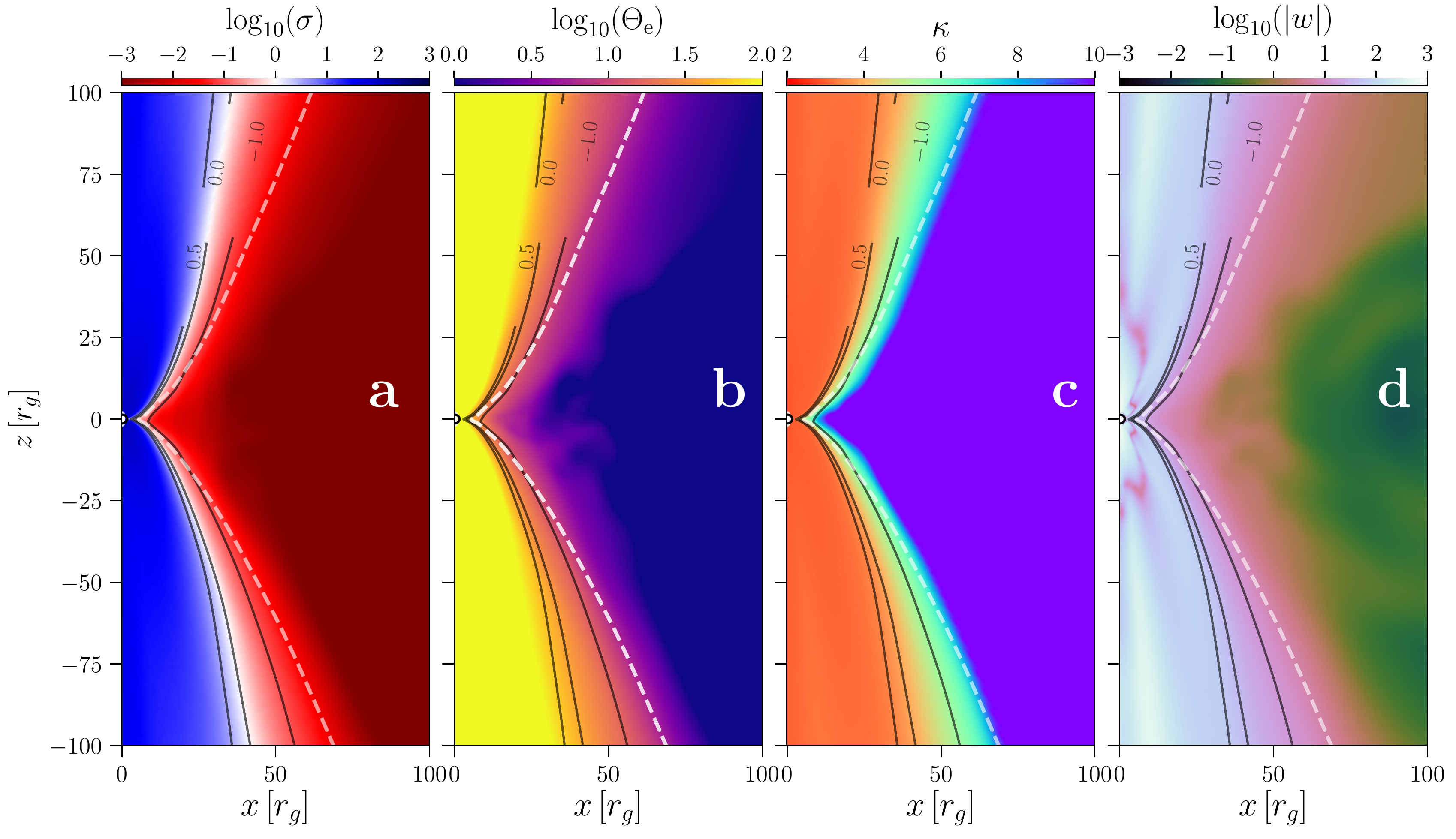}
\caption{\textbf{Small-scale morphology of the jet from GRMHD simulations.} Same as
    in Extended Data Figure 1, Kerr BH with spin $\mathbf{a_{\star}=0.9375}$, but on smaller lengthscales.}
\label{fig:grmhd_short}
\end{center}
\end{figure*}

\begin{figure*}
\begin{center}
\includegraphics[width=0.90\textwidth]{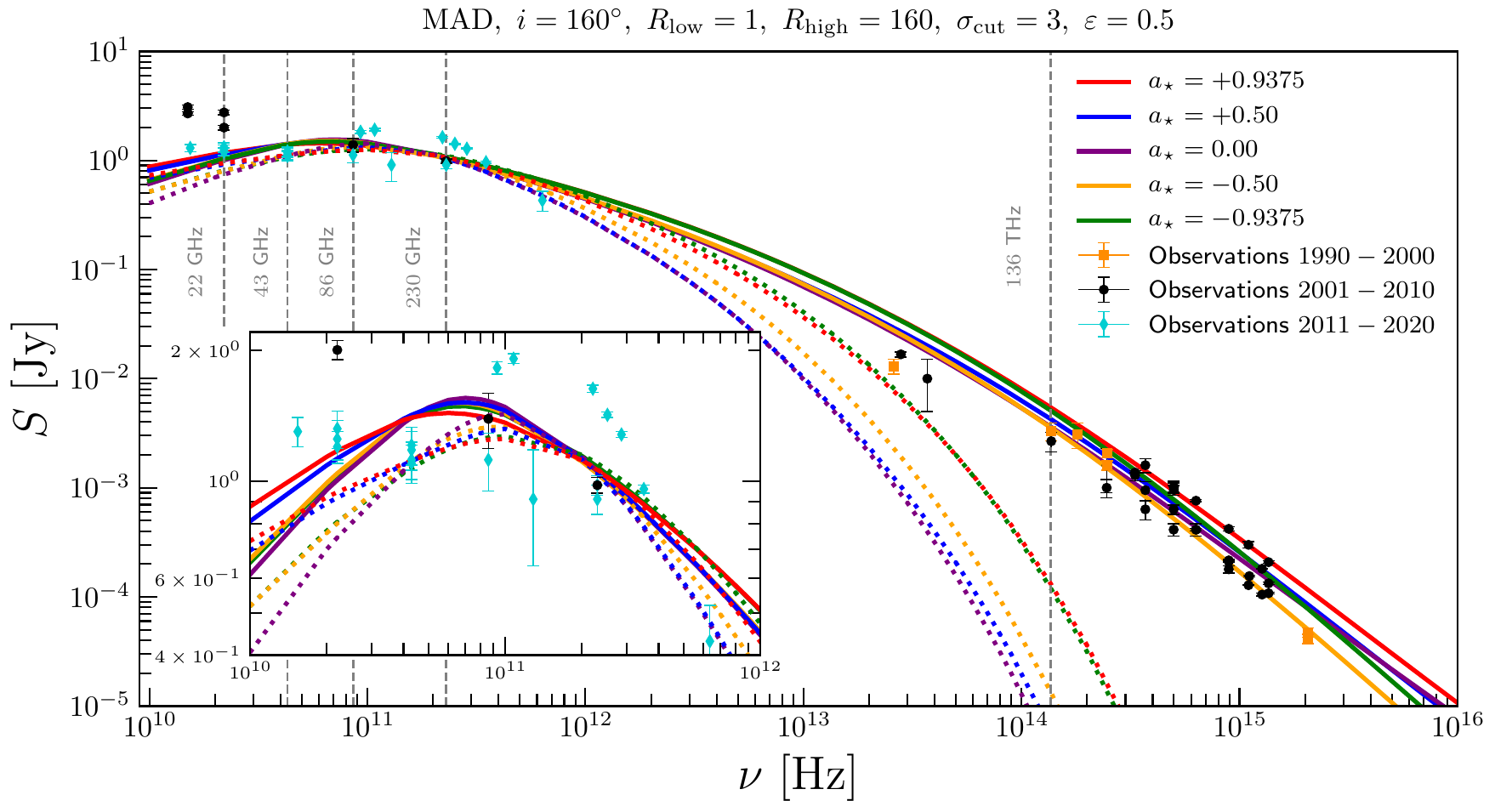}
\caption{\textbf{Same as in Fig. \ref{fig:SEDs},
    but considering five different values of the black hole spin.} Solid
    and dotted lines represent nonthermal and thermal emission models,
    respectively. While gray vertical lines show the most representative 
    frequencies. For each observational data, the uncertainties indicate the 
    variability during the observations.}
\label{fig:sedall}
\end{center}
\end{figure*}

\begin{figure*}
\begin{center}
\includegraphics[width=0.9\textwidth]{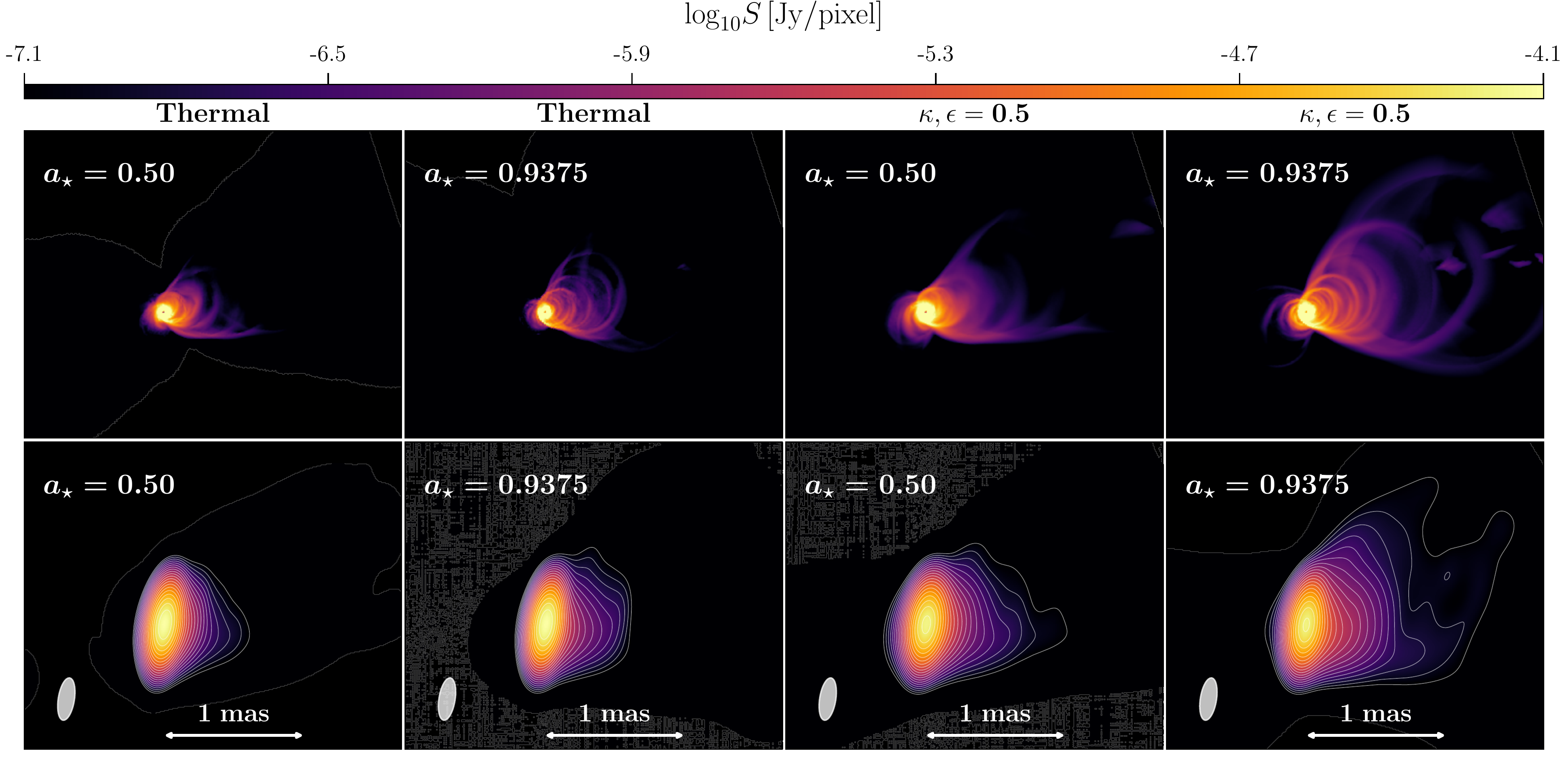}
\caption{\textbf{GRRT and convolved images.} 
The top panels shown the GRRT, while bottom shown convolved 
images for thermal and nonthermal emission models for two
    BHs with spins $\boldsymbol{a_{\star}=0.50}$ and
    $\boldsymbol{a_{\star}=0.9375}$. The images refer to a
  representative time ($t=13,\!820\,M$) and we show as an ellipse in the
  lower-left corner the convolving beam with axes $116\times307\,\mu{\rm
    as}$, as in the observational data. The contours in the flux density
  correspond to $S_{i}=S_{\rm min}+0.43\,{\rm mJy}\times\sqrt{2}^{i}$,
  where $i=0,1,\ldots,n$, such that $S_{n}<S_{\rm max}$.}
\label{fig:teo} 
\end{center}
\end{figure*}
\begin{figure*}
\begin{center}
\includegraphics[width=0.9\textwidth]{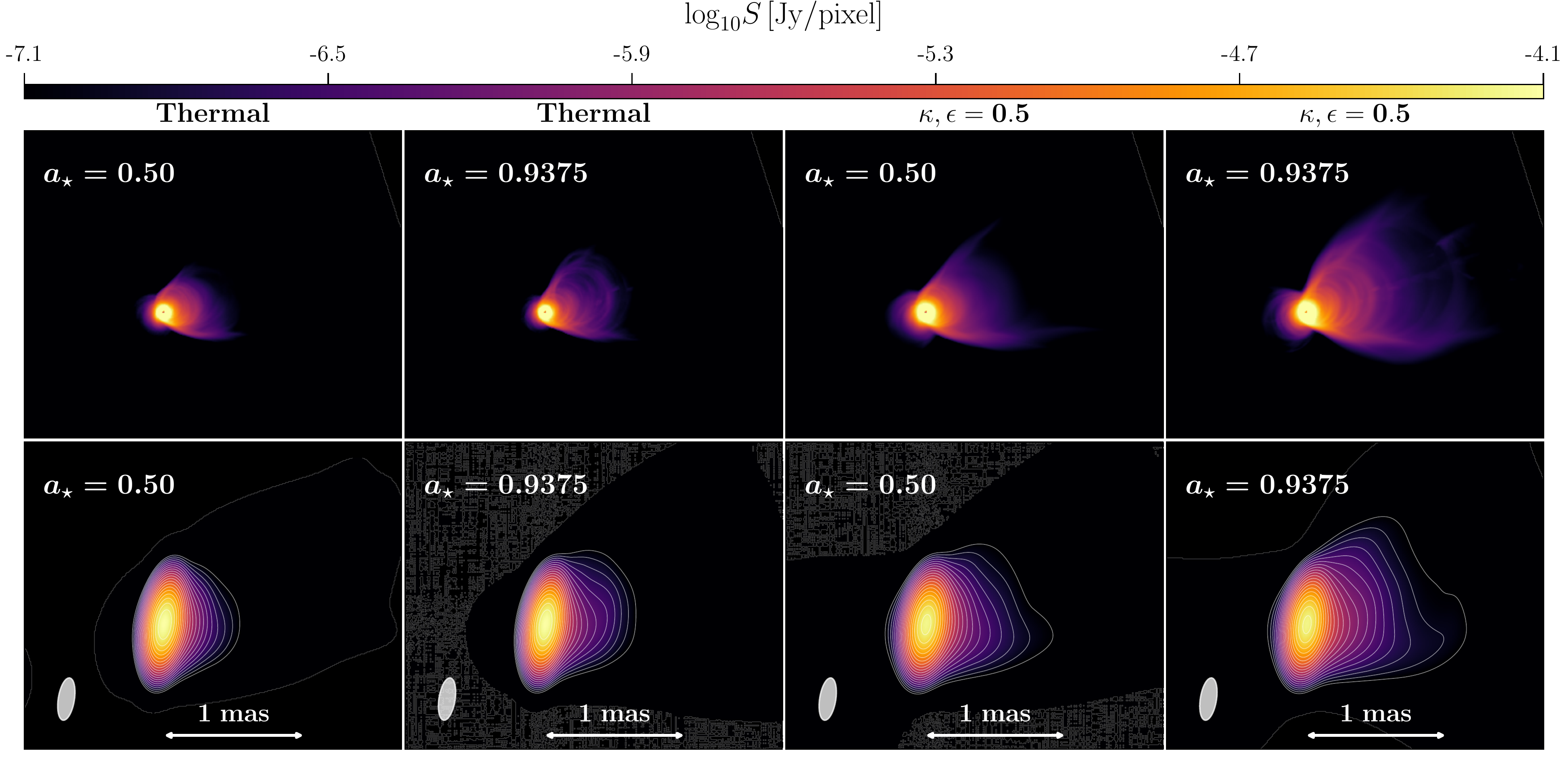}
\caption{\textbf{Same as in Extended Data Figure 4,
    but showing the time average computed between $13,\!000\,M$ and
    $15,\!000\,M$.} Note that all of the features discussed in
    Fig. Extended Data Figure 4 are present also when averaging in time.}
\label{fig:teoave} 
\end{center}
\end{figure*}

\begin{figure*}
\begin{center}
\includegraphics[width=0.3\textwidth]{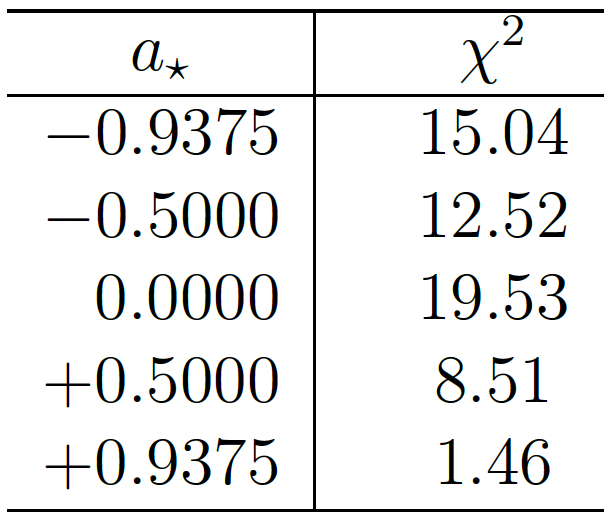}
\caption{\textbf{ $\chi^2$ for the jet width.} $\chi^2$ between the
  observational data and the jet width measured from the numerical
  simulations (see Fig. \ref{fig:jetwidth}).}
\end{center}
\label{tab:imstats}
\end{figure*}

\newpage
\pagebreak
%
%


\end{document}